\documentclass[10pt,sigconf,authorversion,nonacm,letterpaper]{acmart}
\usepackage{graphicx} 
\usepackage{xcolor}
\usepackage{hyperref}
\usepackage{enumitem}
\usepackage{color,soul}
\usepackage{subcaption}
\usepackage{multirow}
\usepackage{enumerate}
\setcopyright{none}

\begin{document}

\title{How Resilient is QUIC to Security and Privacy Attacks?}

\author{Jayasree Sengupta}
\affiliation{%
  \institution{Birla Institute of Technology}
  \country{Mesra, India}}

\author{Debasmita Dey}
\affiliation{%
  \institution{IIEST Shibpur}
  \country{India}
}

\author{Simone Ferlin-Reiter}
\affiliation{%
 \institution{RedHat and Karlstad University}
 \country{Sweden}}

\author{Nirnay Ghosh}
\affiliation{%
  \institution{IIEST Shibpur}
  \country{India}}

\author{Vaibhav Bajpai}
\affiliation{%
  \institution{Hasso Plattner Institute \& University of Potsdam, Germany}
  \country{}}

\begin{abstract}
    QUIC has rapidly evolved into a cornerstone transport protocol for secure, low-latency communications, yet its deployment continues to expose critical security and privacy vulnerabilities, particularly during connection establishment phases and via traffic analysis. This paper systematically revisits a comprehensive set of attacks on QUIC and emerging privacy threats. Building upon these observations, we critically analyze recent IETF mitigation efforts, including TLS Encrypted Client Hello (ECH), Oblivious HTTP (OHTTP) and MASQUE. We analyze how these mechanisms enhance privacy while introducing new operational risks, particularly under adversarial load. Additionally, we discuss emerging challenges posed by post-quantum cryptographic (PQC) handshakes, including handshake expansion and metadata leakage risks. Our analysis highlights ongoing gaps between theoretical defenses and practical deployments, and proposes new research directions focused on adaptive privacy mechanisms. Building on these insights, we propose future directions to ensure long-term security of QUIC and aim to guide its evolution as a robust, privacy-preserving, and resilient transport foundation for the next-generation Internet.

\end{abstract}

\maketitle

\section{Introduction}

QUIC \cite{rfc9000, rfc9369} is a connection-oriented end-to-end encrypted transport protocol based on UDP and built on top of TLS 1.3. Designed to improve upon TCP+TLS by integrating transport and cryptographic handshake layers, QUIC offers features such as reduced connection establishment latency, encrypted transport metadata, and improved resilience to network disruptions. QUIC allows sending parallel and independent data streams which are logically separate from one another, thus ensuring fast and reliable in-order delivery of independent data streams, which avoids the head-of-line blocking problem that TCP typically faces \cite{Chiariotti_COMMAG'21}.

Despite these advancements, QUIC's real-world deployments continue to expose critical security and privacy vulnerabilities. Although QUIC was designed primarily to handle Web traffic, its unique features (such as low-latency, $0-RTT$ benefits) and flexibility (multi-streaming) are suitable for a wide range of applications such as DNS, VPNs and beyond, other than just the Web (HTTP traffic) \cite{Chiariotti_COMMAG'21}. As QUIC's adoption accelerates across web services, content delivery networks, and mobile platforms, systematically understanding and addressing the various security (see: §\ref{sec:attacks}) and privacy (see: §\ref{sec:privacy}) challenges becomes vital-both for protocol designers and for ongoing standardization efforts within the IETF.

Several recent studies have provided valuable overviews of QUIC’s threat landscape. The study \cite{Joardar_TNSM'24} conducted a comprehensive survey of QUIC security and privacy vulnerabilities, threats, and attacks, outlining broad threat categories. Their work proposes high-level research directions but does not deeply examine mitigation efforts underway within the IETF. Meanwhile, Chatzoglou et al. \cite{Chatzoglou'23} revisited QUIC's attack vectors through a detailed review and a hands-on evaluation, focusing on the feasibility of various attacks against early QUIC deployments. However, this study also does not address the implications of newly emerging standards or the operational challenges posed by post-quantum cryptographic integration into QUIC. Thus, while these works underscore the importance of securing QUIC, critical gaps remain. In particular, there is a lack of integrated analysis connecting recent privacy-enhancing mechanisms, their impact and the challenges introduced. Moreover, how new mitigation strategies themselves could create operational vulnerabilities — such as new traffic fingerprinting opportunities — remains underexplored.

In this paper, we address these gaps by systematically revisiting modern attacks on QUIC and its privacy issues. We critically evaluate the security and operational impacts of the emerging IETF mitigation efforts (see: §\ref{sec:mitigation}), identifying risks, and outlining future research directions. Our analysis highlights how stronger privacy protections can inadvertently introduce new operational risks under adversarial conditions, motivating the need for resilience enhancements. We also examine the challenges posed by post-quantum cryptographic integration into QUIC (see: §\ref{sec:quantum}). Through this work, our goal is to contribute a timely and actionable analysis that supports both academic research and practical engineering efforts within the IETF community toward building a more secure, privacy-preserving QUIC transport layer.

\section{Security Attacks on QUIC \label{sec:attacks}}

We begin this section by providing a deep dive into QUIC handshakes, as several security attacks exploit the packets exchanged during QUIC connection establishment. We then discuss the relevant security attacks on QUIC (see: Table \ref{tab:Table1}).

\begin{figure}[!t]
    \centering
    \includegraphics[width=1\linewidth]{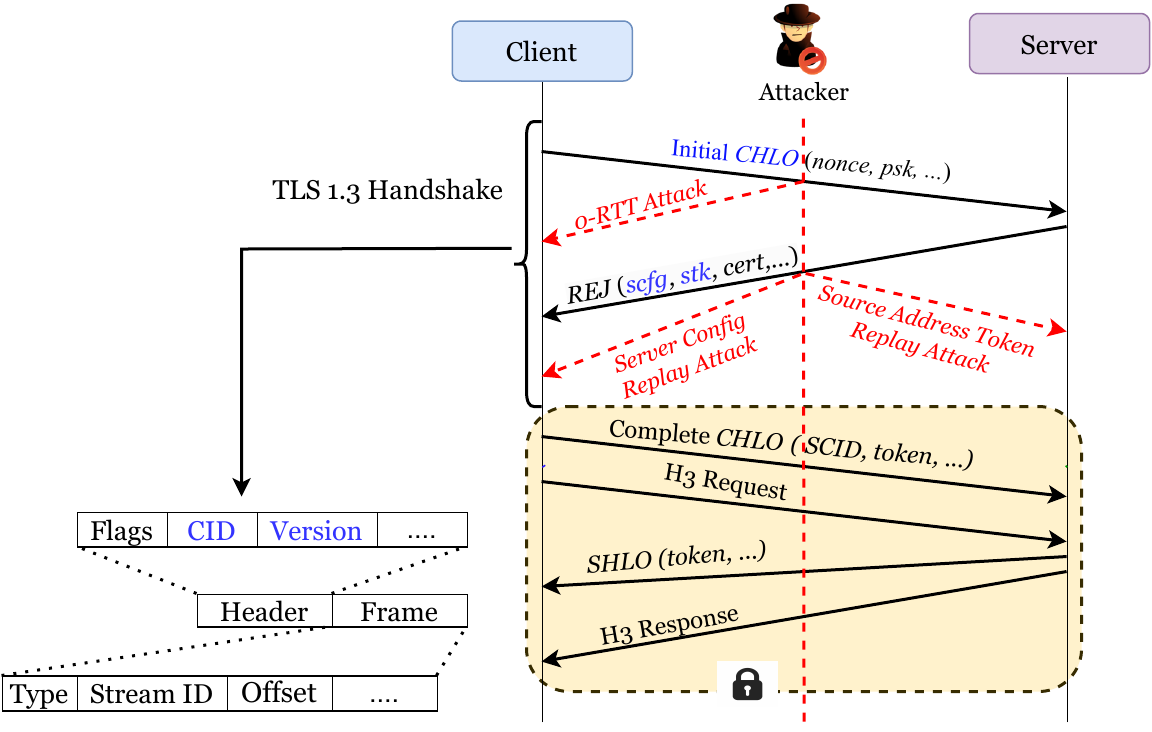}
    \caption{\small \sl QUIC connection establishment with attack vectors. The packets or tuples under attack are represented in blue, whereas the attacks are highlighted in red.}
    \label{fig:1-RTT}
    \vspace{-1em}
\end{figure}

QUIC employs end-to-end encryption, necessitating agreement on multiple parameters between endpoints (client and server). This agreement occurs through a TLS 1.3 handshake, as depicted in Fig. \ref{fig:1-RTT}. A typical QUIC segment's header fields are also illustrated. When a client lacks prior knowledge of a server, it initiates the process by sending an \textit{Initial Client Hello (CHLO)} message to the server. Upon receiving \textit{CHLO}, the server responds with a \textit{reject (REJ)} packet to the client, containing tuples: (i) server config (\texttt{scfg}) with Diffie-Hellman public value, (ii) source-address token (\texttt{stk}) comprising timestamp from the server and client's IP address, (iii) server authentication certificate chain, and (iv) server config signature. After receiving the server config, the client authenticates it using the certificate chain and the server signature. The client then creates a \textit{Complete CHLO}, including its ephemeral Diffie-Hellman key, and sends it back to the server. With the completion of the handshake, the client obtains the initial keys for communication, enabling it to send an encrypted connection request to the server. The server responds with a \textit{server\ hello (SHLO)} packet, encrypted with initial keys and the server's ephemeral keys. Both client and server then calculate their forward secure keys to be used for subsequent communication rounds. The client can now engage in a $0$-RTT handshake, bypassing the initial handshake mechanism, by sending a \textit{Complete\ CHLO}.

\subsection{0-RTT Attack}

During the initiation of a connection setup between the client and server, a $0$-RTT attack can occur when the attacker intercepts the unencrypted initial \textit{CHLO} packet and spoofs it (Fig. \ref{fig:1-RTT}). This opens the door to two potential attacks \cite{0RTT'19}:\\
\noindent\textbf{QUIC RST Attack:} Attacker sends a public reset packet to the client, deceiving it into thinking that the server has rejected the connection. This leads the client to proactively abandon the connection.\\
\noindent\textbf{Version Forgery Attack:} The attacker poses as a server and transmits a version negotiation packet to the client, containing a version unsupported by the client. This prompts the client to either downgrade or abandon the connection.

In this regard, Cao \emph{et al.} \cite{0RTT'19} have detailed the core concepts and implementation of the $0$-RTT attack, emphasizing its potential to result in denial-of-service. They have introduced a mathematical attack description model based on finite state machines to demonstrate the QUIC protocol's vulnerability, showcasing the attack process and confirming QUIC's susceptibility to the described attacks.

\subsection{Replay Attack \label{sec:replay_attack}}

A replay attack occurs when an intruder intercepts a legitimate network transmission and later resends it to deceive the system into treating the re-transmitted data as authentic. Lychev \emph{et al.} \cite{Lychev_SP'15} conducted two types of replay attacks (Fig. \ref{fig:1-RTT}) on the QUIC implementation in Chromium:\\
\noindent\textbf{Server Config Replay Attack:} This occurs when an adversary replays the server's public value  \texttt{(scfg)} to clients who have sent an initial connection request to the server, without being detected by the server. Consequently, these clients generate the initial key and send it to the server, which on being unable to verify their identity, rejects the packets. While the secrecy of the entities involved remains largely unaffected, this attack results in unnecessary consumption of computational resources.\\
\noindent\textbf{Source Address Token Replay Attack:} An intruder replays the source-address token \texttt{(stk)} of a client to the issuing server, allowing the creation of multiple additional connections. This will make the server establish both initial keys and final forward-secure keys. Although subsequent handshake steps would fail, the attacker could potentially launch a denial-of-service (DoS) attack on the server by generating numerous connections on behalf of various clients, leading to a depletion of computational and memory resources.

Fischlin \emph{et al.} \cite{Fischin_EuroSP'17} addressed replay attacks in the $0$-RTT QUIC handshake. While $0$-RTT uses registers for storing \texttt{nonces} to prevent repetition, in QUIC, once the server rejects the $0$-RTT message, the attacker resends the rejected message using a second key to ensure proper delivery. This results in duplicate processing of the same data by the server.

\begin{table*}[!t]
\centering
\caption{Overview of various Security and Privacy Attacks and their Effects on QUIC }
\label{tab:Table1}
\scalebox{0.90}{
\begin{tabular}{|c|c|c|c|c|}
\hline
\textbf{Category} & \textbf{Attack Surface} & \textbf{Attack Name} & \textbf{Effects} & \textbf{Reference} \\ \hline
\multirow{7}{*}{\begin{tabular}[c]{@{}c@{}}Security\\ Attack\end{tabular}} & Initial \textit{CHLO} message & 0-RTT Attack & \begin{tabular}[c]{@{}c@{}}Downgrade; Client-side\\ Connection Rejection\end{tabular} & \cite{0RTT'19} \\ \cline{2-5} 
 & \multirow{2}{*}{\textit{REJ} message} & Replay Attack & \begin{tabular}[c]{@{}c@{}}Connection Failure;\\ Server DoS\end{tabular} & \cite{Lychev_SP'15,Fischin_EuroSP'17} \\ \cline{3-5} 
 &  & \begin{tabular}[c]{@{}c@{}}Source-Address Token (\texttt{stk})\\ Manipulation Attack\end{tabular} & \multirow{2}{*}{\begin{tabular}[c]{@{}c@{}}Connection Failure;\\ Server Load\end{tabular}} & \cite{Lychev_SP'15} \\ \cline{2-3} \cline{5-5} 
 & \multirow{2}{*}{Packet Header} & \begin{tabular}[c]{@{}c@{}}Connection ID (\texttt{cid})\\ Manipulation Attack\end{tabular} &  & \cite{Lychev_SP'15} \\ \cline{3-5} 
 &  & Request Forgery Attack & Traffic Amplification & \cite{Gbur_NDSS'23} \\ \cline{2-5} 
 & \begin{tabular}[c]{@{}c@{}}Handshake and\\ Packet Header\end{tabular} & \begin{tabular}[c]{@{}c@{}}State-Overflow and\\ Resource Exhaustion Attacks\end{tabular} & \begin{tabular}[c]{@{}c@{}}Fake connections;\\ Resource Wastage\end{tabular} & \cite{Nawrocki_IMC'21} \\ \cline{2-5} 
 & QUIC Server & QUIC Flood DDoS Attack & \begin{tabular}[c]{@{}c@{}}Resource Exhaustion;\\ System breakdown\end{tabular} & \cite{Nawrocki_IMC'21,DDoS_Cloudflare} \\ \hline
\multirow{2}{*}{\begin{tabular}[c]{@{}c@{}}Privacy\\ Attack\end{tabular}} & \multirow{2}{*}{\begin{tabular}[c]{@{}c@{}}Traffic traces and\\ Communication Channel\end{tabular}} & User Tracking via QUIC & \begin{tabular}[c]{@{}c@{}}User Profiling;\\ Targetted Advertising\end{tabular} & \cite{Sy_PoPETS'19} \\ \cline{3-5} 
 &  & Website Fingerprinting & User Privacy breach & \cite{Zhan_CN'21,Smith_USENIX'22} \\ \hline
\end{tabular}%
}
\end{table*}

\subsection{Manipulation Attack}

A manipulation attack aims to compromise the key agreement process, leading the client and server to agree on distinct keys. It is achieved by altering unprotected packet fields, such as connection id (\texttt{cid}) or the source-address token (\texttt{stk}) (Fig. \ref{fig:1-RTT}) that are utilized as input to the key derivation process. Lychev \emph{et al.} \cite{Lychev_SP'15} describe the following two types of manipulation attacks in QUIC:

\noindent\textbf{Connection ID Manipulation Attack:} Occurs when an attacker generates a fresh \texttt{cid}, causing the server and client to perceive different \texttt{cid} values. Though the handshake begins normally, the server and the client have distinct encryption keys due to different \texttt{cid} values. Consequently, decryption fails, and failed packets are buffered until the handshake is completed. In due time, this situation leads to the disconnection of the link. However, the error message gets encoded using the original encryption key, thereby preventing its decryption and preserving the state until it times out.\\
\noindent\textbf{Source-Address Token Manipulation Attack:} In this attack, the attacker generates separate \texttt{stk} values for both the client and server, resembling a CID manipulation attack. The handshake typically begins with the transmission of \textit{CHLO} messages. However, since encryption keys rely on \texttt{stk} as an input, the client and server generate distinct encryption keys, leading to decryption failure. As a result, the client buffers failed \textit{CHLO} messages and retransmits them for 10 seconds. After a timeout, the client sends an encrypted error message to the server. Yet, as the server possesses initial encryption keys, the connection remains intact, and this state may persist for 10 minutes.

\subsection{UDP Hole Punching Bypass Attack}

Another attack scenario that may arise due to QUIC's inherent characteristics is the UDP hole punching bypass attack. QUIC, commonly used on top of UDP, is known to be compatible with stateful firewalls. UDP  typically require an examination of individual packets to establish a connection. However, stateful firewalls do not provide this level of verification for UDP connections, which leaves them vulnerable. In \cite{arXiv_Gbur'21}, Konrad \emph{et al.} highlights that a server can be compromised by an attacker who uses remote code execution. Owing to the connectionless feature of UDP, any 5-tuple message from the server side may be captured to create a hole for the consequent packets with similar 5 tuples. This allows the attacker to keep the connection active by sending packets at regular intervals. As a result, the connection may remain open for a significant period of time, even after it has been terminated.

\subsection{Request Forgery Attack}

This attack involves instigating a client to send deceptive requests to another client. In \cite{Gbur_NDSS'23}, Gbur \emph{et al.} assume a scenario where the attacker has complete control over packets delivered to the victim, enabling three types of forgery attacks:\\
\noindent\textbf{Server Initial Request Forgery (SIRF):} An attacker initiates a QUIC handshake with the server, by spoofing the packet's source IP address and port fields. This tricks the victim to assume that the connection is from a genuine server.\\
\noindent\textbf{Version Negotiation Request Forgery (VNRF):} In this attack, the attacker sends an unknown version in the client's packet, prompting the QUIC server to respond with a version negotiation packet. This action triggers the version negotiation functionality.\\
\noindent\textbf{Connection Migration Request Forgery (CMRF):} While QUIC's connection migration feature offers advantages, a drawback exists: the server is unable to differentiate between a genuinely migrated client address and a spoofed one. In this scenario, an attacker conducts a legitimate handshake with the server to establish a new connection, then spoofs an arbitrary packet to obtain the source address and sends it to the server. Consequently, the server, upon detecting a new source address, unwittingly establishes the connection and transmits UDP packets to the manipulated address.

\subsection{State-Overflow and Resource Exhaustion Attacks \label{sec:exhaustion}}

In these attacks, adversaries masquerade as legitimate clients and initiate full handshakes with QUIC servers. The server responds with a unique Source Connection ID (\texttt{SCID}) and the corresponding TLS certificate, thus allocating resources to maintain connection states. To overwhelm the server, the adversarial client randomly spoofs source ports and IP addresses, flooding the server with multiple handshake requests. This results in the creation of numerous unique \texttt{SCIDs} and concurrent connection states, leading to server overload, state overflow, and resource exhaustion — similar to \texttt{TCP\ SYN} floods, where legitimate client requests may also be rejected. The QUIC design's primary vulnerability lies in the lack of client verification during the initial round-trip of the QUIC full handshake.

A study \cite{Nawrocki_IMC'21} evaluates these attacks on real-world QUIC traffic based on active measurements using the UCSD network telescope. Results show that, similar to \texttt{TCP SYN} floods, the QUIC handshake is susceptible to resource exhaustion attacks. Notably, 98\% of these attacks target QUIC servers of well-known companies, with Google being the primary target (58\%) and Facebook accounting for 25\% of the attacks. While QUIC originally supports \textit{RETRY} messages to mitigate resource exhaustion attacks, their implementation adds an extra round-trip time (RTT), conflicting with QUIC's claimed performance gains. This happens as a \textit{RETRY} message typically precedes a QUIC handshake and forces the client to prove its authenticity by responding with a unique token. The study \cite{Nawrocki_IMC'21} emphasizes that during experimentation, no \textit{RETRY} messages were observed, indicating a lack of practical defense mechanisms in QUIC deployments.

In order to limit the effect of resource exhaustion attacks, QUIC originally supports \textit{RETRY} messages. A \textit{RETRY} message typically precedes a QUIC handshake and forces the client to prove its authenticity by responding with a unique token. However, this adds another RTT which eventually conflicts with the performance gains claimed by QUIC. In this regard, the study \cite{Nawrocki_IMC'21} further reveals that they did not observe any \textit{RETRY} messages during their experimentation which reconfirms the lack of practical defense mechanisms in QUIC deployments.

\subsection{QUIC Flood DDoS Attack \label{sec:ddos}}

As previously explained, a QUIC flood DDoS attack occurs when an attacker overwhelms a QUIC server with a large volume of data, causing the victim server to slow down significantly or crash. Defending against DDoS attacks in QUIC is challenging because QUIC is based on UDP, which provides minimal or no information for blocking illegitimate traffic. Additionally, as QUIC packets are encrypted, the victim server cannot easily verify the legitimacy of the data source. While some Content Distribution Networks (CDNs) like \textit{Cloudflare} have been successful in mitigating QUIC floods \cite{DDoS_Cloudflare}, a study in \cite{Nawrocki_IMC'21} indicates that these attacks persist in the real world. In~\cite{Nawrocki_IMC'21}, the authors reported that the Internet is vulnerable to four QUIC flood attacks every hour. Among these, 51\% of the attacks occur concurrently with TCP/ICMP floods, while another 40\% target the victim sequentially. However, QUIC floods have a shorter duration, lasting only $255\ seconds$, compared to their TCP/ICMP counterparts. However, such shorter duration may lead to faster resource exhaustion or system breakdown, presenting an aspect that is yet to be explored.

\section{Privacy Attacks on QUIC \label{sec:privacy}}
This section reviews two specific privacy threats that are posed to QUIC-based user applications.  

\subsection{User Tracking via QUIC \label{sec:user_tracking}}

Online tracking threatens user privacy by exploiting users' browsing habits, which may reveal sensitive information. It can then be used for profiling, web analytics, and targeted ads. To counter this, browsers need robust privacy protection. While QUIC, being encrypted is widely adopted across Web browsers as it aims to ensure user privacy, it remains susceptible to tracking through the following two mechanisms:\\
\noindent\textbf{Linking several website visits by the same user:} The \textit{source-address-token} is a unique data block included in the \textit{reject} (REJ) message sent by the server to a client during initial connection setup. The client caches it and presents to the server for every new $0$-RTT connection setup with the same server. This enables the server to identify subsequent connection requests from the same user which enables linking different website visits to the same usernames.\\
\noindent\textbf{User tracking across multiple sessions:} QUIC clients cache \textit{server-config} (part of REJ message), containing a unique 16-byte \textit{server config identifier} (SCID) assigned by the server to each user. This SCID allows Web servers and potential attackers to link the initial connection request to subsequent requests using the same SCID within \texttt{CHLO} messages.

The study \cite{Sy_PoPETS'19} reveals QUIC's privacy violation, enabling trackers to map users based on tokens provided during a $0$-RTT connection setup attempt. Analysis of popular browsers (Chrome) revealed insufficient protective measures against Web user tracking. Such monitoring was particularly effective in resource-constrained scenarios due to lower bandwidth requirements and delays compared to HTTP cookies or traditional browser fingerprinting.

\subsection{Website Fingerprinting \label{sec:wf}}

An encrypted transmission protocol is susceptible to \textit{website fingerprinting (WFP)} attacks if adversaries can deduce a user's visited websites by monitoring the transmission channel. As QUIC encrypts data, adversaries target unencrypted handshake packets. Further, as QUIC is based on a request-response model, it allows attackers to distinguish traffic from different Web resources due to the relatively fixed browser rendering sequence \cite{Zhan_CN'21}. This weakens privacy benefits achieved through encryption under WFP attacks.

Zhan \emph{et al.} \cite{Zhan_CN'21} explored WFP attack vulnerabilities in both QUIC variants: (i) \textit{GQUIC} (Google’s QUIC) and (ii) \textit{IQUIC} (IETF’s QUIC). \textit{GQUIC} is more vulnerable in restricted traffic scenarios, while both protocols exhibit similar vulnerabilities under typical traffic. Multiple WFP attacks with only 40 packets achieved 95.4\% and 95.5\% accuracy for \textit{GQUIC} and \textit{IQUIC}, respectively.

To counter WFP attacks, \textit{IQUIC} introduces a \texttt{PADDING} frame, but studies suggest network-level padding inefficiency against adversaries capable of observing traffic traces. Current research \cite{Smith_USENIX'22} focuses on integrating defenses directly into client applications (e.g., browsers) by effectively using the \texttt{PADDING} frame to thwart adversary's attempt of traffic analysis in Virtual Private Networks (VPNs).

A summary of various security and privacy attacks on QUIC is presented in Table \ref{tab:Table1}.

\section{Potential Mitigation Strategies \label{sec:mitigation}}

The previous two sections have broadly highlighted the emerging security and privacy challenges, particularly centered around metadata leakage and handshake vulnerabilities. As QUIC matures into the de facto transport protocol for modern secure communications, addressing its evolving security and privacy vulnerabilities has become a critical research and operational priority. This section critically evaluates emerging mitigation strategies standardized or proposed within the IETF, outlines unresolved challenges, and identifies promising research directions to make QUIC more resilient for future Internet deployments.

\subsection{Encrypting Handshake Metadata}

Despite TLS1.3 being encrypted, certain information are exchanged in plaintext during the initial connection establishment phase, allowing the eavesdroppers to learn various crucial metadata and other sensitive information like Server Name Identification \texttt{(SNI)} extension (see: § \ref{sec:attacks}) available in the \textit{CHLO} message. This drawback particularly motivated the IETF to encrypt the handshake messages apart from just the application data. IETF proposes the inclusion of the \textit{Encrypted Client Hello (ECH)} extension \cite{ietf-tls-esni-24} into the TLS to mitigate the security risks and improve the privacy of the Internet as a whole. The goal of \textit{ECH} is to protect all sensitive handshake parameters such as preshared key, server name, etc. by encrypting the entire \textit{CHLO}, thereby closing the gap left by TLS 1.3 and \textit{Encrypted Server Name Identification (ESNI)}, the predecessor of \textit{ECH}. A recent study \cite{Bhargavan_CCS'22}, shows that \textit{ECH}, indeed ensures basic security goals (like confidentiality, authenticity etc.), while offering stronger downgrade resistance handshake privacy and broader metadata protection. Despite these benefits offered by ECH, several practical challenges still persist:

\begin{itemize}[leftmargin=*]
    \item \textbf{Partial Deployment and Downgrade Risks:} Many servers still do not support ECH, leading to fallback to unencrypted handshakes. Attackers can exploit this partial deployment by forcing fallback, exposing sensitive metadata.
    \item \textbf{Middlebox Interference:} Numerous middleboxes incorrectly assume visibility into TLS extensions. RFC 9325 \cite{rfc9325} outlines middlebox tolerance recommendations, but real-world compliance remains limited.
    \item \textbf{Traffic Analysis on Encrypted Handshakes:} A recent study \cite{Trevisan_TOIT'23} reveals that simple machine learning models are capable of recovering a the domain names of a user's visited websites even in the presence of \textit{ECH} with very high accuracy. Attackers can infer the target server or user activity through traffic features such as handshake size, packet timing, and record counts, even when ECH is employed. Even in the presence of standard padding-based techniques, attackers are still able to gain a sufficiently large amount of information
    \item \textbf{Dependency on DNS Security:} ECH relies on the correct and secure delivery of public keys via DNS SVCB/HTTPS \cite{rfc9460} records. Attacks on DNS privacy, such as via DNS cache poisoning or compromised resolvers, undermine ECH’s guarantees.
\end{itemize}

Future research should prioritize developing efficient, traffic-aware padding and shaping mechanisms to better protect handshake messages against traffic analysis attacks. In parallel, it is crucial to enforce integrated defenses that span both DNS resolution and TLS handshakes, ensuring that metadata privacy is preserved throughout the resolution and connection establishment phases; specifically, tight coupling between DNS-over-QUIC (DoQ) \cite{rfc9250} and TLS encryption is necessary to prevent potential leakage through resolution paths. Additionally, improving the deployability of \textit{ECH} remains a pressing concern, motivating the exploration of progressive deployment strategies such as enabling clients to attempt encrypting \textit{CHLO} messages even in the absence of explicit server-side \textit{ECH} advertisements.

\subsection{Privacy-preserving Proxying}

The IETF has developed complementary protocols to strengthen user privacy beyond the transport layer. Oblivious HTTP (OHTTP) (RFC 9484 \cite{rfc9458}) decouples client IP addresses from request payloads by relaying encrypted HTTP requests through trusted intermediaries. This design ensures that no single entity can observe both the sender’s identity and the content of the request. Similarly, MASQUE (Multiplexed Application Substrate over QUIC Encryption) \cite{schinazi-masque-proxy-05} enables tunneling of arbitrary IP and UDP flows over QUIC connections, thereby supporting privacy-preserving proxying, virtual private network (VPN) services, and flexible secure tunneling for broader Internet applications.

Despite these advances, OHTTP and MASQUE introduce several operational and security challenges. First, OHTTP assumes that relays and target gateways do not collude; however, in adversarial environments, relay compromise or collusion can severely undermine user privacy guarantees. Second, both OHTTP and MASQUE proxies may be exploited for amplification attacks unless they implement robust rate-limiting, authentication, and authorization mechanisms to control resource usage. Third, additional relaying inevitably introduces latency and bandwidth overheads, raising trade-offs between enhanced privacy and acceptable performance metrics, particularly for real-time or mobile applications.

\subsection{Resilience Against Resource Exhaustion}

While mechanisms such as \textit{ECH} and privacy-preserving proxying via OHTTP and MASQUE strengthen user privacy, they also increase server-side computational and memory demands during connection establishment. As outlined in §~\ref{sec:exhaustion} and ~\ref{sec:ddos}, this heightened state maintenance makes QUIC endpoints more vulnerable to resource exhaustion and DoS attacks, where adversaries exploit expensive handshake operations to overwhelm servers.

To mitigate these risks without negating the privacy benefits of ECH and OHTTP, several approaches can be adopted. Refining QUIC’s stateless \textit{RETRY} mechanisms can enable earlier validation of client authenticity during handshake attempts, ensuring that malicious clients are filtered with minimal server resource commitment. Further, under extreme load conditions, servers may employ graceful privacy degradation strategies, such as offering opportunistic encryption modes that maintain basic confidentiality without the full overhead of \textit{ECH} or multi-hop proxying, thus ensuring continued service availability without complete privacy forfeiture. Together, these approaches ensure that privacy enhancements do not inadvertently compromise the operational resilience of QUIC deployments.

\section{QUIC in the Post-Quantum Era \label{sec:quantum}}

While the mitigation strategies discussed in §~\ref{sec:mitigation} address current threats to QUIC security and privacy, an important future challenge is ensuring resilience against quantum-capable adversaries. Cryptographically relevant quantum computers could compromise classical public key systems, weakening handshake security and long-term confidentiality. To proactively mitigate this risk, the IETF is standardizing hybrid key exchange approaches that combine classical and post-quantum algorithms \cite{ietf-tls-hybrid-design-12}, while also developing broader deployment guidelines for transport and application-layer protocols \cite{ietf-pquip-pqc-engineers-09, reddy-uta-pqc-app-07}. 

Integrating post-quantum cryptography (PQC) into QUIC presents several operational concerns. Post-quantum key encapsulation mechanisms (KEMs) notably increase handshake sizes, exacerbating risks of resource exhaustion attacks (§~\ref{sec:exhaustion},~\ref{sec:ddos}) by amplifying computational and memory burdens on servers. Additionally, hybrid handshakes may create new traffic analysis vectors (§~\ref{sec:wf}), allowing adversaries to distinguish handshake types based on packet sizes, flow timings, or retransmission patterns. Such metadata leakage could undermine the privacy goals of mechanisms like \textit{ECH}, especially if hybrid and classical flows become easily identifiable.

Mitigating these risks requires compact hybrid handshake designs that minimize handshake expansion while preserving cryptographic strength. Furthermore, formal modeling of handshake privacy under hybrid and post-quantum settings is necessary to evaluate indistinguishability guarantees. Finally, traffic shaping and padding mechanisms must evolve to equalize observable characteristics across classical, hybrid, and future post-quantum-only sessions, preventing adversarial profiling. Ensuring that QUIC remains resilient in a post-quantum world will demand careful balancing of cryptographic security, metadata privacy, and operational scalability. Addressing these engineering challenges proactively is essential to maintaining QUIC's role as a secure, future-proof transport protocol.

\section{Conclusion}

This paper revisited the evolving security and privacy challenges faced by QUIC, analyzing vulnerabilities in handshake metadata exposure, traffic analysis, and resilience under adversarial conditions. We critically evaluated emerging mitigation strategies, including encrypted handshake mechanisms, privacy-preserving proxy architectures, and resource exhaustion defenses, highlighting both the advances made and the challenges that persist in real-world deployments. Additionally, we examined the operational complexities introduced by post-quantum cryptographic handshakes, identifying new risks related to handshake amplification and metadata leakage. Our findings reveal that while significant progress has been made, important deployment gaps persist despite theoretical advancements. Bridging this gap will require continued research into adaptive privacy mechanisms, traffic pattern obfuscation, cross-layer resilience, and post-quantum readiness. Proactive engineering towards robust, privacy-preserving, and quantum-resilient QUIC is critical to securing the next-generation Internet.

\bibliographystyle{ACM-Reference-Format}
\bibliography{references}
\end{document}